\documentclass[floatfix,aps,prb,twocolumn,showpacs,superscriptaddress]{revtex4}
\usepackage{graphicx}
\usepackage[]{epsfig}
\usepackage{times,amsmath,amssymb}
\begin{document}
\title{Probing local electronic states in the quantum Hall regime with a side coupled quantum dot}

\author{Tomohiro Otsuka}
\email[]{t-otsuka@issp.u-tokyo.ac.jp}
\affiliation{Institute for Solid State Physics, University of Tokyo, 5-1-5 Kashiwanoha, Kashiwa, Chiba 277-8581, Japan}

\author{Eisuke Abe}%
\affiliation{Institute for Solid State Physics, University of Tokyo, 5-1-5 Kashiwanoha, Kashiwa, Chiba 277-8581, Japan}

\author{Yasuhiro Iye}%
\affiliation{Institute for Solid State Physics, University of Tokyo, 5-1-5 Kashiwanoha, Kashiwa, Chiba 277-8581, Japan}

\author{Shingo Katsumoto}%
\affiliation{Institute for Solid State Physics, University of Tokyo, 5-1-5 Kashiwanoha, Kashiwa, Chiba 277-8581, Japan}
\affiliation{Institute for Nano Quantum Information Electronics, University of Tokyo, 4-6-1 Komaba, Meguro, Tokyo 153-8505, Japan}

\date{\today}
\begin{abstract}
We demonstrate a new method for locally probing the electronic states in the quantum Hall regime
utilizing a side coupled quantum dot positioned at an edge of a Hall bar.
By measuring the tunneling of electrons from the Hall bar into the dot,
we acquire information on the local electrochemical potential and electron temperature.
Furthermore, this method allows us to observe the spatial modulation of the electrostatic potential at the edge state due to many-body screening effect.
\end{abstract}

\maketitle


The edge states~\cite{1988ButtikerPRB} formed in two dimensional electron gases (2DEG's) under strong magnetic fields play important roles in the transport properties in the quantum Hall (QH) effect~\cite{1980KlitzingPRL}. 
They are formed as a consequence of the Landau quantization and the confinement potential at the edges of the devices.
In conventional experimental methods for the study of electronic states in the QH effect, gvoltage probesh, which are composed of macroscopic Ohmic contacts, are used.
Although the voltage difference between the contacts can be measured, this is not enough to explore microscopic properties of the edge states.
Also this macroscopic contacts induce relaxation of the electronic states in the contacts and it is inevitable to change the original electronic states in the QH effect.

For exploring the microscopic electronic states, local probes utilizing liquid helium films~\cite{1991KlassZPhys}, the Pockels effect~\cite{1992FronteinSrfSci}, cyclotron emission~\cite{1999KawanoPRB}, and scanning probe microscopy~\cite{1998TessmerNat,1999YacobySSC,2000FinkelsteinSci,2008HashimotoPRL} have been reported.
They succeeded to show real space images of electric fields or electron densities.
Nevertheless local electrostatic and thermodynamic properties are still elusive.
In this study we apply side coupled quantum dots (QD's) to obtain local electrochemical potential, electron temperature and spatial configuration of the edge states. Since it is easy to obtain a side coupled QD in a the few electron regime with keeping tunneling probability to the edge~\cite{2007OtsukaJPSJ}, we can use a well defined single level in the QD and this enables high energy resolution.
Also the flow of electrons between the edge and the QD is regulated by the Coulomb blockade and can be very small (less than fA) and the measurement has very small disturbance to the original electronic states.
Though the positions of the QD's are fixed, the high energy resolution and the sensitivity to QD-edge distance enable us to detect characteristic variation of the electrostatic potential in the QH effect~\cite{1992ChklovskiiPRB,1994LierPRB}.

 
\begin{figure}
\begin{center}
  \includegraphics{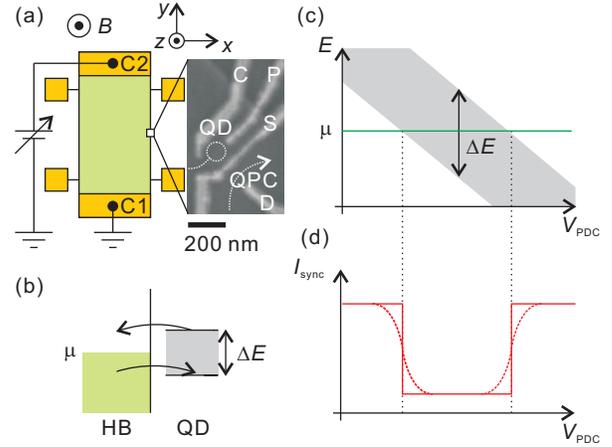}
  \caption{(color online) (a) Schematic of Device~1.
The scanning electron micrograph shows the QD part of the device.
(b) Energy diagram when the electron shuttling occurs.
(c) Shift of the energy window (the gray zone) as a function of $V_{\rm PDC}$.
(d) $I_{\rm sync}$ as a function of $V_{\rm PDC}$. The solid (broken) line shows an expected result at low (high) electron temperature $T_{\rm e}$.}
  \label{Scheme}
\end{center}
\end{figure}

We measured two devices fabricated from a GaAs/AlGaAs heterostructure wafer
with the sheet carrier density of 2.1~$\times$~10$^{15}$~m$^{-2}$ and the mobility of 32~m$^2$/Vs.
After the formation of Ohmic contacts,
36~$\mu$m~$\times$~108~$\mu$m-sized Hall bars (HB's) were patterned by wet-etching,
followed by the deposition of Au/Ti Schottky gates to define QD's.
The size of the HB's is sufficiently large for observing the QH effect.
As depicted in Fig.~\ref{Scheme}(a), one of them (Device~1) has a QD in the middle of the right edge.
The other device (Device~2, not shown) has an additional QD placed 6~$\mu $m apart from contact C1 on the right edge.
The devices were cooled with a dilution refrigerator (base temperature around 30~mK), and the perpendicular magnetic field $B$ was applied using a superconducting solenoid.

Here we measure the local electronic states utilizing a QD side coupled to the edge state.
In the following we briefly summarize the technique, which is fully described in Refs.~\onlinecite{2004ElzermanAPL,2008OtsukaAPL,2009OtsukaPRB}.
A QD and the edge state are separated by a potential barrier formed by a Schottky 
gate and we detect tunneling events between them through changes in the number of electrons in the QD.
This detection is realized by a remote charge detector utilizing a quantum point contact (QPC) placed next to the QD~\cite{1993FieldPRL,1998BuksNat,2003ElzermanPRB}.

By applying square wave voltages on gate P $V_{\rm P}$, we regularly shift the chemical potential of the QD and form an energy window with the width $\Delta E$ [Fig.~\ref{Scheme}(b)].
When the electrochemical potential $\mu$ in the HB is in this window, the potential shift causes electron shuttling between the edge and the QD, and this causes synchronous current modulation through the QPC.
Actually direct electrostatic coupling between gate P and the QPC also leads to a background synchronous current modulation and the effect of the electron shuttling appears as a decrease of the synchronous current $I_{\rm sync}$.
As sweeping the DC offset voltage on gate P $V_{\rm PDC}$ and continuously shifting the energy window [Fig.~\ref{Scheme}(c)],
a dip of $I_{\rm sync}$ is observed in the region in which $\mu $ and the energy window cross [Fig.~\ref{Scheme}(d)].
This dip structure contains information on the local electronic states in the vicinity of the tunneling barrier of the QD.



\begin{figure}
\begin{center}
  \includegraphics{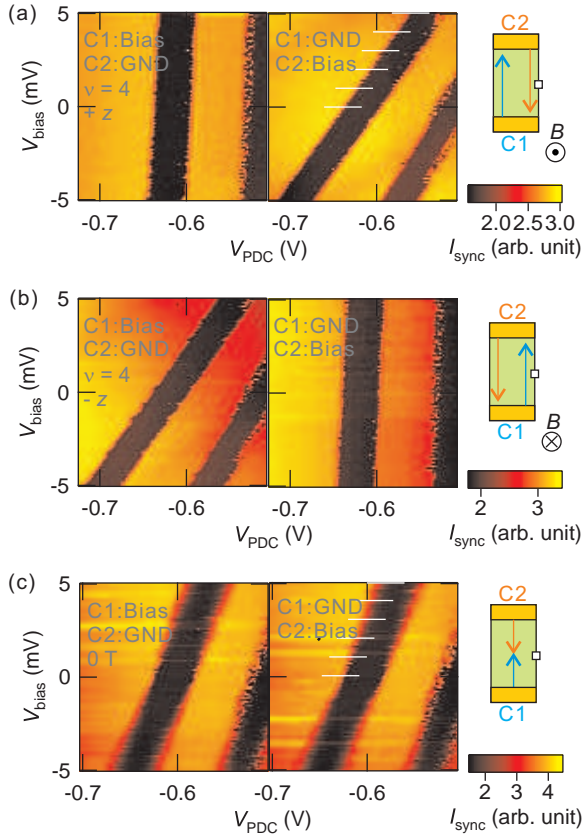}
  \caption{(color online) $I_{\rm sync}$ as a function of $V_{\rm PDC}$ and $V_{\rm bias}$ in the positive (a), negative (b) and zero (c) magnetic fields.
The filling factor is $\nu$ = 4 in both (a) and (b).
The left (right) graphs are the results when $V_{\rm bias}$ is applied on C1 (C2).
The schematics in the right hand side illustrate the direction of the edge states. The horizontal lines in (a) and (c) correspond to the data in Fig.~\ref{EleTmp}.}
  \label{ElePot}
\end{center}
\end{figure}

Figure~\ref{ElePot} shows the observed $I_{\rm sync}$ of Device~1 as a function of $V_{\rm PDC}$ and the bias voltage on the HB $V_{\rm bias}$.
The bias was applied on one contact while the other contact was connected to the ground.
The left (right) graphs show the results when $V_{\rm bias}$ is applied on contact C1 (C2).
The number of electrons in the QD is set to zero or one.
In the QH regime with the filling factor $\nu$ = 4 [Fig.~\ref{ElePot}(a)],
the dip positions (bands in dark color) are almost fixed when C1 is biased, while linear shifts are observed when C2 is biased.
Note that the survival of the QH effect in this bias range is confirmed by measuring the longitudinal resistance with conventional voltage probes.

This asymmetry in the bias condition is reasonable considering that the system is in the QH regime, where the voltage drop along $y$ direction is zero except at the hot spots~\cite{1991KlassZPhys,1999KawanoPRB} near the ejection contacts.
The result is also viewed as a consequence of the chirality of the edge states.
When the magnetic field is applied in $+z$ direction [Fig. \ref{ElePot}(a)], the electrons emitted from C2 enter the right edge state.
Then the electrochemical potential of the right edge state is same to that of C2.
When the direction of the magnetic field is reversed, the electrochemical potential of the right edge follows C1.
The results shown in Fig. \ref{ElePot}(a) and (b) are in good agreement
with the above deduction.
We attribute the small shifts of the dip positions in the left graph of Fig. \ref{ElePot}(a) and the right graph of Fig. \ref{ElePot}(b) to the contact resistances,
which induce small voltage changes at the contacts.

At zero magnetic field, the electrochemical potential varies linearly along $y$ direction between the two contacts irrespective of the bias condition because the edge states or other mechanisms which suppress the energy relaxation are not available.
Since the QD is positioned halfway between the contacts, the shift of the dip positions is just the half of that in the QH regime as observed in Fig.~\ref{ElePot}(c).


\begin{figure}
\begin{center}
  \includegraphics{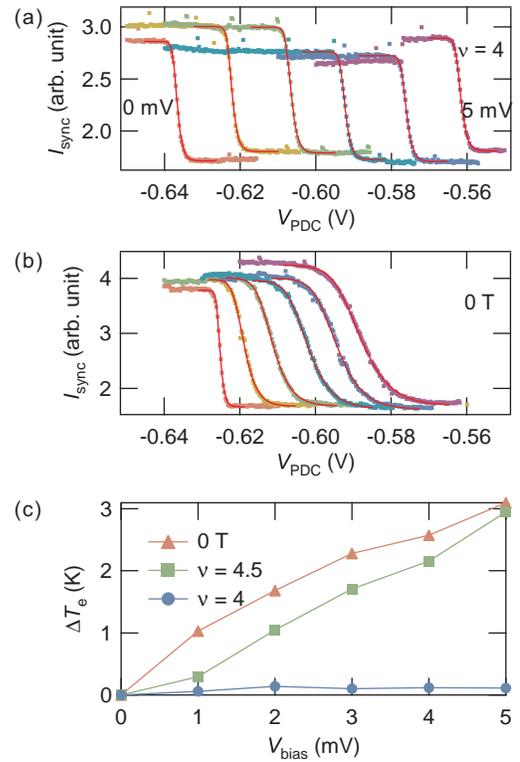}
  \caption{(color online) $I_{\rm sync}$ as a function of $V_{\rm PDC}$ at $\nu$ = 4 (a) and in the zero magnetic field (b).
The solid lines superposed on the data are the results of the fitting using $T_{\rm e}$ as a fitting parameter.
From the left to the right, $V_{\rm bias}$ was varied from 0 to 5~mV by 1~mV. (c) $\Delta T_{\rm e}$ as a function of $V_{\rm bias}$.}
  \label{EleTmp}
\end{center}
\end{figure}

We now focus on the line shape of the dip in Fig.~\ref{ElePot}.
It is observed in Fig.~\ref{ElePot}(c) that the boundaries of the dip are progressively blurred as $V_{\rm bias}$ becomes larger.
On the contrary, the boundaries are always sharp in the QH regime [Fig.~\ref{ElePot}(a) and (b)].
As the dip occurs when $\mu $ and the energy window intersect, 
its sharpness reflects the sharpness of the electron distribution around $\mu $ and thus the local electron temperature $T_{\rm e}$, as illustrated in Fig.~\ref{Scheme}(d).
The cross sections along the white lines in Fig.~\ref{ElePot} are shown in Fig.~\ref{EleTmp}.

For quantitative evaluation, we assume that the broadening follows the Fermi-Dirac distribution
\begin{equation}
\label{FD}
F(E) = [\exp \left\{ (E-\mu )/k_{\rm B}T_{\rm e} \right\}+1]^{-1},
\end{equation}
where $k_{\rm B}$ is the Boltzmann constant.
To apply Eq.~(\ref{FD}), the coefficient $\alpha$ to convert $V_{\rm PDC}$ into the energy $E$ is necessary.
By modeling a current-carrying channel as a series circuit of the zero-resistance one-way conductor ({\it i.e.}, edge channel) and resistors at the two contacts,
we obtain $\alpha = (1/t_1+1/t_2)^{-1}$, where $t_{1}$ and $t_{2}$ are the tangents of the dips when C1 and C2 are biased, respectively~\cite{remark1}.
Therefore, $\alpha$ is directly obtained from Fig.~\ref{ElePot}.
The solid lines in Fig.~\ref{EleTmp}(a) and (b) show the results of the fitting using $T_{\rm e}$, an additional offset and a magnitude factor as fitting parameters.
For comparison, we also evaluate $T_{\rm e}$ at $\nu$ = 4.5.
At zero bias, we obtain $T_{\rm e}$ = 523, 621 and 856~mK for $B$ = 0~T, $\nu$ = 4 and $\nu$ = 4.5, respectively.
At this stage, it is not certain what causes the high $T_{\rm e}$ and the differences between them.
One possible reason is the radiation of the noise from the QPC~\cite{2006OnacPRL,2007GustavssonPRL,2008HashisakaPRB}.
But it is possible to analyze the effect induced by $V_{\rm bias}$ because the effect is large and we can extract that by evaluating the increase of the electron temperature $\Delta T_{\rm e}(V_{\rm bias}) = T_{\rm e}(V_{\rm bias}) - T_{\rm e}(V_{\rm bias} = 0)$.
$\Delta T_{\rm e}$ is plotted in Fig.~\ref{EleTmp}(c).
As increasing $V_{\rm bias}$, $\Delta T_{\rm e}$ in non-QH conditions becomes large up to as high as 3~K.
On the contrary, $\Delta T_{\rm e}$ at $\nu$ = 4 is nearly zero, regardless of $V_{\rm bias}$.
This certifies the lack of the scattering mechanisms that raise $T_{\rm e}$ in the QH regime.


\begin{figure}
\begin{center}
  \includegraphics{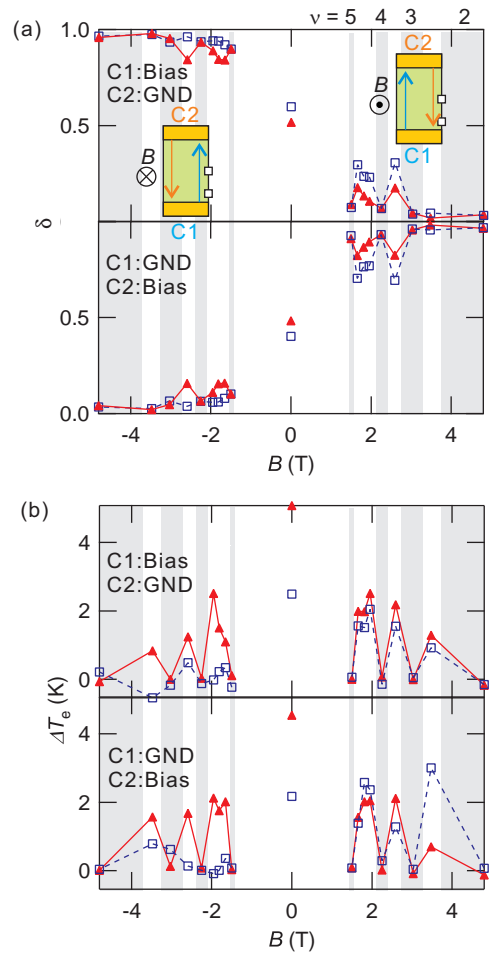}
  \caption{(color online) (a) $\delta \equiv \Delta \mu /\Delta V_{\rm bias}$ as a function of $B$~\cite{remark2}.
The triangles (squares) are the results of the QD at the center (near C1).
The upper (lower) graph shows the result when C1 (C2) is biased.
The gray regions are the QH regimes with $\nu$ as indicated.
The insets show schematics of the edge states in negative and positive magnetic fields.
(b) $\Delta T_{\rm e}$ as a function of $B$. The symbols are the same as (a).
}
  \label{PosDep}
\end{center}
\end{figure}

Next we measure Device~2 in order to examine the position- and $B$-dependence of $\mu $ and $\Delta T_{\rm e}$.
Figure~\ref{PosDep}(a) and (b) respectively show the change of $\mu$ with the change of $V_{\rm bias}$, $\delta \equiv \Delta \mu /\Delta V_{\rm bias}$,
and $\Delta T_{\rm e}$ as a function of $B$ at $V_{\rm bias}$ = 4~mV at the two QD positions.
In the QH regimes (gray regions), the values of $\delta$ are very close to zero or one,
depending on whether the corresponding edge state is in equilibrium with the grounded or biased contact.
Also, $\Delta T_{\rm e}$ is very small, as expected.
It is consistent with the nature of the edge states that these features do not depend on the position along the device edge.
In non-QH regimes, we observe that $\delta$ behaves in a manner similar to that in the QH regimes
although the values are not as close to zero or one as the latter.
Furthermore, $\delta$ at the middle position is symmetric with regard to $B$,
while $\delta$ near C1 is asymmetric (closer to zero or one in negative magnetic fields).
In the transition region between the QH regimes suppression of backscattering and energy relaxation is lifted.
This deviates $\delta $ from zero or one.
As for the asymmetry, it is explained as a result of the difference in the degree of the energy relaxation.
In negative fields, the electrons from C1 enter the QD near C1 without suffering the energy relaxation.
Thus, the values of $\delta$ are very close to zero or one.
On the other hand, in positive fields, the electrons from C2 enter the QD after large energy relaxation.
In much the same reason, the asymmetry is also observed in $\Delta T_{\rm e}$  for the QD near C1.

Note that the benefit of this method with a side coupled QD is the small disturbance~\cite{2009OtsukaPRB} to the original electronic states.
It is different from the measurement with conventional voltage probes, in which relaxation in the probes is inevitable.
With this property, it becomes possible to measure the degree of the relaxation shown in Fig. 4.


So far, we have confirmed that our method is capable of probing basic features of the electronic states in the QH regime,
such as the chirality, and the absence of energy relaxation.
We now proceed to obtain more detailed information on the edge states, namely, the spatial modulation of the electrostatic potential.
\begin{figure}
\begin{center}
  \includegraphics{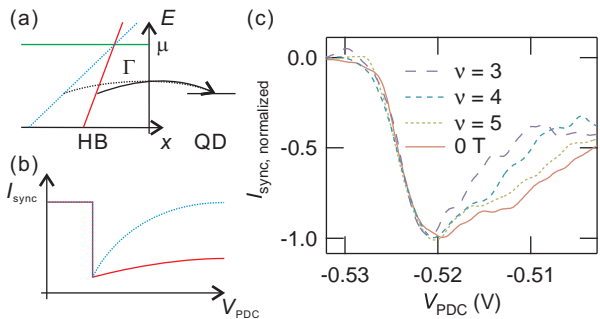}
  \caption{(color online) (a) Energy diagram near the device edge.
The solid (broken) line shows the Landau level without (with) screening around $\mu$.
(b) $I_{\rm sync}$ as a function of $V_{\rm PDC}$ for cases without (solid line) and with (broken line) screening.
(c) Normalized $I_{\rm sync}$ as a function of $V_{\rm PDC}$.}
  \label{Screen}
\end{center}
\end{figure}
In the theory beyond the single-particle picture,
the reconstruction of the electrostatic potential leads to the formation of stepwise distributions (along $x$) of the edge states at absolute zero temperature~\cite{1992ChklovskiiPRB}.
Even at finite temperature, this screening effect survives and makes the gradient of the Landau level ${\rm d}E/{\rm d}x$ near $\mu $ smaller than that without screening [Fig.~\ref{Screen}(a)]~\cite{1994LierPRB}.
While the signatures of the screening effect have been observed in experiments on inter-edge tunneling~\cite{1996MachidaPRB} and AB type oscillation in antidots~\cite{2009KatoPRL},
the present method gives more direct and detailed access to this effect.

The idea is utilizing the relation between the gradient ${\rm d}E/{\rm d}x$ and the tunneling rate $\Gamma$.
Since $\Gamma$ depends exponentially on the tunneling distance, 
it is highly sensitive to the change of ${\rm d}E/{\rm d}x$ [arrows in Fig.~\ref{Screen}(a)].
The dip depth $\Delta I_{\rm sync}$ and $\Gamma $ are related by the formula~\cite{2004ElzermanAPL}
\begin{equation}
\Delta I_{\rm sync}\propto1-\frac{\pi^2}{\Gamma^2 /4f^2+\pi^2},
\label{eq_depth}
\end{equation}
where $f $ is the frequency of the square wave.
By adjusting $f$ to be comparable with $\Gamma$, we can realize the condition in which $\Delta I_{\rm sync}$ is sensitive to $\Gamma$ and consequently to ${\rm d}E/{\rm d}x$.
Note that in the preceding measurements the condition $\Gamma /2f\gg 1$ was employed and $\Delta I_{\rm sync}$ was nearly constant.
Here, the bottom of the dip is no longer flat, but rather shows a buildup structure with the increase of $V_{\rm PDC}$ reflecting the barrier thickness, as illustrated in Fig.~\ref{Screen}(b).
It is expected that the stronger screening results in a faster buildup of the bottom of the dip because of the faster decrease of $\Gamma$. 
In this way, we are able to investigate the electronic states below $\mu $.

Figure~\ref{Screen}(c) shows the observed $I_{\rm sync}$ as a function of $V_{\rm PDC}$.
The traces show the results at $\nu$ = 3, 4, 5 as well as $B$ = 0~T.
$I_{\rm sync}$ is normalized to make the deepest points equal to $-1$.
In the measurement, we set the experimental conditions for $\Gamma$ to be equal at the respective deepest points.
This procedure is to compensate the possible change in the QD-edge state distance at the Fermi energy as $B$ is varied and to extract the pure change induced by the modification of ${\rm d}E/{\rm d}x$.  
It is observed that the buildup of $\Delta I_{\rm sync}$ is faster at smaller $\nu$.
This implies the stronger screening at smaller $\nu$.
In our method, we are probing only the outermost channel because it has by far the largest tunneling probability among the channels.
The degeneracy in this channel becomes larger at higher fields.
This could enhance the electron-electron interactions and facilitate the redistribution of the electrons.
This interpretation qualitatively explains the observed phenomena.


In conclusion, we have investigated the local electronic states in the quantum Hall regime utilizing side coupled quantum dots as local probes.
We have observed the formation of the edge states, and confirmed their chirality.
We have succeeded in determining the local electron temperature, and confirmed the suppression of energy relaxation in quantum Hall regime.
Finally, we have investigated the screening  effect in the edge states.
Our results demonstrate the ability of the new method to deduce the local information on the quantum Hall states,
which is not obtained through the conventional transport measurements.
This method will be applicable to approach the hotspots and edge states in the fractional quantum Hall effect.  

{\it Note added}: After completion of our work, we became aware of a paper by Altimiras {\it et al}. about non-equilibrium edge-channel spectroscopy utilizing a quantum dot between two edge channels~\cite{2009AltimirasNatPhys}.

We thank A. Endo, M. Kato and T. Fujisawa for fruitful discussions, and Y. Hashimoto for technical supports.
This work was supported by Grant-in-Aid for Scientific Research and Special Coordination Funds for Promoting Science and Technology.


\begin{references}
\bibitem{remark1}
This formula is unchanged when a finite resistance is introduced in the conductor, and thus valid even in non-QH conditions.
\bibitem{remark2}
Although the QH effect was also observed in lower fields below $|B|$ = 1.5~T, 
it appears that opening of the Landau gaps in the QH regime is not large enough to allow for clear comparison with non-QH regimes.
At fields above $|B|$ = 4.5~T, the QD's were not stable.
Thus, thorough measurements were not carried out in those ranges of $B$.
\end{references}
\end{document}